\documentstyle[floats,aps,epsf,amsfonts]{revtex}
\begin{document}

\title{Regularization parameters for the self force in Schwarzschild
spacetime:\\ II. gravitational and electromagnetic cases}
\author{Leor Barack$^1$ and Amos Ori$^2$}
\address{
$^1$Albert-Einstein-Institut, Max-Planck-Institut f{\"u}r
Gravitationsphysik, Am M\"uhlenberg 1, D-14476 Golm, Germany.
$^2$Department of Physics, Technion---Israel Institute of Technology,
Haifa, 32000, Israel.\\
}
\date{\today}
\maketitle


\begin{abstract}

We obtain all ``regularization parameters'' (RP) needed for calculating
the gravitational and electromagnetic self forces for an arbitrary geodesic
orbit around a Schwarzschild black hole.
These RP values are required for implementing the previously introduced
mode-sum method, which allows a practical calculation
of the self force by summing over contributions from individual multipole modes
of the particle's field. In the gravitational case, we provide
here full details of the analytic method and results briefly reported in a
recent Letter [Phys.\ Rev.\ Lett.\ {\bf 88}, 091101 (2002)]. In the
electromagnetic case, the RP are obtained here for the first time.

\end{abstract}


\section{introduction} \label{SecI}

This is the second in a series of papers aimed to establish a practical
calculation scheme for the self force 
acting on a point particle in orbit around a black hole. This scheme---referred
to as the ``mode-sum method''---stems from the general
regularization prescription by Mino, Sasaki, and Tanaka (MST)\cite{MST} and
Quinn and Wald (QW)\cite{QW}. In effect, the mode-sum method re-formulates the
MST-QW general result in the language of multipole modes, thereby making
it accessible to standard numerical treatment. In practice, the application
of this method involves two basic parts: (i) calculation of the ``full''
modes of the force, through numerical integration of the decoupled field
equations; and (ii) analytical derivation of certain parameters (whose values
depend on the orbit under consideration) called the ``regularization parameters''
(RP). Previously, the explicit values of the RP were derived analytically
in a few special cases of orbits in Schwarzschild spacetime---specifically,
for circular and radial orbits in the scalar field case \cite{MSRS-scalar},
and for radial trajectories in the gravitational case \cite{MSRS-grav,BL-radial}.
In these works, the RP values where calculated through a rather cumbersome
local expansion of the ($l$-multipole) Green's function. The application of
this technique to more general orbits appears a challenging task.

In a recent Letter \cite{Letter}, the joint groups of Barack and Ori (BO)
and Mino, Nakano, and Sasaki (MNS) devised an alternative, more direct
method for obtaining the RP. The new method is based on a multipole
decomposition of the explicit ``direct'' part of the force (see below).
Using this method, BO and MNS were able to calculate the explicit RP
values for both the scalar and gravitational self forces, for any geodesic
orbit in Schwarzschild spacetime. In a preceding paper \cite{paperI}
(hereafter referred to as ``paper I'') BO described the full
details of the new calculation technique, as applied to the toy-model
of the {\em scalar} self force acting on a scalar charge.
The current paper deals with the more interesting case of the
{\em gravitational} self force on a mass particle, and provides full
details of the RP derivation in this case.
In addition, we shall
derive here the RP values for the {\em electromagnetic} self force acting
on an electrically charged particle orbiting a Schwarzschild black hole.
(BO and MNS applied two slightly different methods in obtaining
the RP; MNS describe their calculation in \cite{MNS}.)

The analysis presented in this paper relies greatly on the technique
and results of Paper I, to which we shall frequently refer the reader.
Though the basic idea of our calculation is the same as in the scalar
toy model of Paper I, some unavoidable technical complexities arise
when coming to consider the gravitational or electromagnetic cases. In
particular, one then has to consider an extension of the particle's four-velocity
(which takes part in constructing the ``direct force''---see below) off
the worldline, and address the question of the RP dependence
on the (non-unique) choice of such extension. Another, more fundamental
issue, is the gauge
dependence of the gravitational self force and its implication to the
mode sum scheme (see Ref.\ \cite{gauge}).

Most of our manuscript will concern with the (most interesting)
gravitational case. Our analytical technique is easily applicable
to the (simpler) electromagnetic case, which we shall later consider
in a separate section. The structure of this paper is as follows:
In the rest of this introductory section we set up the physical scenario
of a pointlike mass particle orbiting a Schwarzschild black hole,
introduce MST-QW's prescription for calculating the gravitational
self force on this particle, and review the basics of the mode-sum method.
In Sec.\ \ref{SecII} we present MST's expression for the direct
part of the gravitational self force, and analytically process this
expression to extract the information relevant for deriving the RP.
Section \ref{SecIII} contains the heart of our
calculation, namely, the derivation of all RP for any geodesic orbit,
in the gravitational case. Section \ref{SecV} deals with the electromagnetic
case. In Sec.\ \ref{SecVI} we summarize the prescription for calculating
the gravitational and electromagnetic self forces via the mode-sum method,
and give some concluding remarks.

Throughout this paper we use geometrized units
(with $G=c=1$), metric signature ${-}{+}{+}{+}$,
and the standard Schwarzschild coordinates $t,r,\theta,\varphi$.

\subsection{Pointlike particle model}

We consider a pointlike particle of mass $\mu$, moving freely
in the vacuum exterior of a Schwarzschild black hole with mass $M\gg\mu$.
(QW \cite{QW} discuss the extent to which the concept of a pointlike
particle makes sense in the context of the radiation reaction problem.)
In the limit $\mu\to 0$, the particle traces a geodesic $z^{\mu}(\tau)$
of the Schwarzschild background.
Due to angular momentum conservation, the geodesic orbit (as well as
the orbit under self-force effect) is confined to a plain, which, without
loss of generality, we shall take as the equatorial plain,
$\theta=\pi/2$.

When the mass $\mu$ is finite, the particle no longer moves on a geodesic.
In this case, it is useful to view the particle
as being subject to a self force induced by its own
gravitational field (treated as a perturbation over the background
geometry).\footnote {One might attempt an alternative point of
view, which regards the particle as freely moving in a ``perturbed
spacetime''. This picture, though, is somewhat problematic, as the
perturbed spacetime is singular at the point-particle limit, and hence
is not strictly defined at the particle's location.
See, however, Ref.\ \cite{DW}.}
The particle's equation of motion thus takes the form
\begin{equation}\label{I-20}
\mu u^{\alpha}_{\ ;\beta}u^{\beta}=F^{\alpha}_{\rm self},
\end{equation}
where $u^{\alpha}\equiv dz^{\alpha}/d\tau$,
a semicolon denotes covariant differentiation with respect
to the background geometry, and $F^{\alpha}_{\rm self}\propto O(\mu^2)$
describes the leading-order self-force effect.
In the following we shall consider the self force acting on
the particle at an arbitrary point along its worldline,
denoted by $z=z_0\equiv (t_0,r_0,\theta_0,\varphi_0)$
(in our setup $\theta_0=\pi/2$).
We shall use the notation $x\equiv (t,r,\theta,\varphi)$ to
represent a point in the neighborhood of $z_0$.

We will denote the metric of the perturbed spacetime as
$g_{\alpha\beta}+h_{\alpha\beta}$, where $g_{\alpha\beta}$ is the (Schwarzschild)
background metric and $h_{\alpha\beta}(\propto\mu)$ is the metric perturbation
induced by the particle. Following MST-QW, we consider the metric perturbation
$h_{\alpha\beta}$ specifically in the harmonic gauge.
We shall denote by $\bar{h}_{\alpha\beta}$ the trace-reversed perturbation:
\begin{equation}\label{I-10}
\bar{h}_{\alpha\beta}\equiv h_{\alpha\beta}-\frac{1}{2}g_{\alpha\beta}.
\end{equation}

\subsection{Gravitational self force according to MST-QW}

MST and QW found that the gravitational self force on a particle freely
moving in a vacuum spacetime can be formally constructed as \cite{DW}
\begin{equation}\label{I-30}
F^{\alpha}_{\rm self}=\lim_{x\to z_0} F^{\alpha}_{\rm tail}(x),
\end{equation}
where $F^{\alpha}_{\rm tail}$ is the ``tail'' force, associated with
the mere effect of waves scattered {\em inside} (rather than propagating along)
the particle's past light cone.
The tail force can be derived from the ``tail'' part of the
metric perturbation, as defined by MST \cite{MST}, through
\begin{equation} \label{I-33}
F^{\alpha}_{\rm tail}(x)=\mu\, k^{\alpha\beta\gamma\delta}(x)
\bar{h}^{\rm tail}_{\beta\gamma;\delta}(x).
\end{equation}
Here, the tensor $k^{\alpha\beta\gamma\delta}(x)$ is any
(sufficiently regular) extension of the tensor
\begin{eqnarray} \label{I-50}
k_0^{\alpha\beta\gamma\delta}=\frac{1}{2}g^{\alpha\delta}u^{\beta}u^{\gamma}
-g^{\alpha\beta}u^{\gamma}u^{\delta}-\frac{1}{2}u^{\alpha}u^{\beta}
u^{\gamma}u^{\delta}+\frac{1}{4}u^{\alpha}g^{\beta\gamma}u^{\delta}
+\frac{1}{4}g^{\alpha\delta}g^{\beta\gamma},
\end{eqnarray}
defined at $x=z_0$,
where $u^{\alpha}$ and $g^{\alpha\delta}$ refer to the values of the
four-velocity and the metric tensor at $z_0$.
[Later we shall employ a specific extension of
$k_0^{\alpha\beta\gamma\delta}$; note that the choice of extension
does not affect the physical self force
$F^{\alpha}_{\rm self}$, though, obviously, it does affect the
field $F^{\alpha}_{\rm tail}(x)$ off the worldline.]

The (singular) difference between the ``full'' perturbation
$\bar h_{\alpha\beta}(x)$
and the tail part $\bar h^{\rm tail}_{\alpha\beta}(x)$ is associated with
the instantaneous effect of waves propagating directly along the particle's
light cone. This part is referred to as the ``direct'' perturbation:
\begin{equation}\label{I-55}
\bar h^{\rm dir}_{\alpha\beta}(x)\equiv
\bar h_{\alpha\beta}(x)-\bar h^{\rm tail}_{\alpha\beta}(x).
\end{equation}
Correspondingly, we define the ``direct'' force as
\begin{equation} \label{I-75}
F^{\alpha}_{\rm dir}(x)\equiv \mu\, k^{\alpha\beta\gamma\delta}(x)
\bar{h}^{\rm dir}_{\beta\gamma;\delta}(x).
\end{equation}
Defining also the ``full'' force,
\begin{equation} \label{I-70}
F^{\alpha}_{\rm full}(x)\equiv \mu\, k^{\alpha\beta\gamma\delta}(x)
\bar{h}_{\beta\gamma;\delta}(x),
\end{equation}
we then have
\begin{equation}\label{I-60}
F^{\alpha}_{\rm tail}(x)=
F^{\alpha}_{\rm full}(x)-F^{\alpha}_{\rm dir}(x).
\end{equation}
The explicit form of the direct perturbation has been derived
by MST \cite{MST} (see also \cite{MN,MNS}). It is given below in Eq.\
(\ref{II-10}), and serves as the starting point for our analysis.

Note that both the direct force and the full force, which were defined
above as vector fields in the neighborhood of $z_0$, diverge as $x\to z_0$.
However, their difference, yielding the tail force, admits a perfectly
regular limit $x\to z_0$, which, according to MST-QW, represents the
physical self force. In this respect, notice also the freedom one has
in choosing the extension $k^{\alpha\beta\gamma\delta}(x)$,
as long as this extension is regular enough and reduces to
$k^{\alpha\beta\gamma\delta}_0$ in the limit $x\to z_0$.
One has to make sure, though, that {\em the same
extension $k^{\alpha\beta\gamma\delta}(x)$} is applied to both the
direct and the full forces.


\subsection{Mode-sum method}

The mode-sum method was previously introduced
\cite{MSRS-scalar,MSRS-grav} as a practical method for
calculating the MST-QW self force given in Eq.\ (\ref{I-30})
\cite{Lousto,implementation}.
The method is reviewed in Paper I; here we merely describe
the basic prescription (as applied to the gravitational case) and
introduce the relevant notation.

In the mode-sum scheme, one first formally expands the gravitational
tail force, as well as the full and direct forces, into multipole
$l$-modes, in the form
\begin{equation}\label{I-80}
F^{\alpha}_{\rm tail,full,dir}(x)=
\sum_{l=0}^{\infty}F^{\alpha l}_{\rm tail,full,dir}(x).
\end{equation}
Here, precisely as in the scalar case, the modes $F^{\alpha l}_{\rm tail}$,
$F^{\alpha l}_{\rm full}$, and $F^{\alpha l}_{\rm dir}$ are obtained by
decomposing (each of the vectorial components of) the corresponding quantities
$F^{\alpha}_{\rm tail}$,
$F^{\alpha}_{\rm full}$, and $F^{\alpha}_{\rm dir}$ into standard scalar
spherical harmonics, and then, for any given multipole number $l$, summing
over all azimuthal numbers $m$. It is important to emphasize here that
the various $l$-modes introduced in Eq.\ (\ref{I-80}) are defined in our
scheme through a {\em scalar} harmonic decomposition. In this regard, recall
that the (full) metric perturbation in Schwarzschild spacetime is usually
decomposed into {\em tensor}
harmonic modes in actual calculations. The construction of the full-force
scalar-harmonic modes $F^{\alpha l}_{\rm full}$ from the full perturbation
tensor-harmonic modes can be prescribed in a straightforward
manner (as, e.g., in \cite{BL-radial,BL-circular}).

The basic prescription for constructing the gravitational self force
via the mode-sum scheme is given by \cite{MSRS-scalar,MSRS-grav}
\begin{equation}\label{I-90}
F^{\alpha}_{\rm self}=\sum_{l=0}^{\infty}\left[\lim_{x\to z_0}
F^{\alpha l}_{\rm full}(x)-A^{\alpha}L-B^{\alpha}-C^{\alpha}/L\right]
-D^{\alpha},
\end{equation}
where $L\equiv l+1/2$ and the ($l$-independent) coefficients
$A^{\alpha}$, $B^{\alpha}$, $C^{\alpha}$, and $D^{\alpha}$ are
the {\em regularization parameters} (RP).
The RP $A^{\alpha}$, $B^{\alpha}$, and $C^{\alpha}$ may be defined
by the demand that the sum in Eq.\ (\ref{I-90}) would converge.
Equivalently (and more practically), one may define these parameters
by requiring convergence of the sum
\begin{equation}\label{I-100}
\sum_{l=0}^{\infty}\left[\lim_{x\to z_0}F^{\alpha l}_{\rm dir}(x)
-A^{\alpha}L-B^{\alpha}-C^{\alpha}/L\right]\equiv D^{\alpha}.
\end{equation}
This sum then defines the fourth parameter, $D^{\alpha}$.
From the above definitions it is clear that the RP values may be derived
through analysis of the direct force modes $F^{\alpha l}_{\rm dir}(x)$.

Eq.\ (\ref{I-90}) constitutes a practical prescription for constructing
the gravitational self force, given (i) the values of all necessary RP,
and (ii) the full-force modes $F^{\alpha l}_{\rm full}$.
In this paper we derive all RP for any (equatorial) geodesic orbit in
Schwarzschild spacetime, hence setting an analytical basis for
calculations of the gravitational self force for all such orbits.

\section{Analyzing the direct gravitational force}\label{SecII}

\subsection{Direct part of the metric perturbation}

\begin{figure}[thb]
\input{epsf}
\centerline{\epsfysize 6cm \epsfbox{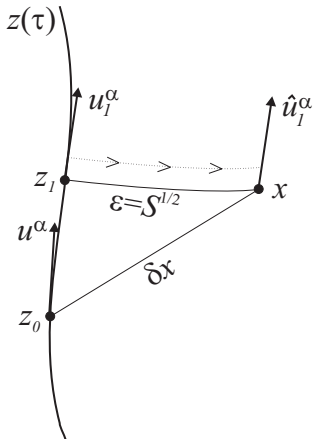}}
\caption{\protect\footnotesize
Geometric setup and notation: The self force is calculated at the
point $z=z_0$ on the geodesic $z(\tau)$. $x$ is an off-worldline
point in the neighborhood of $z_0$, and $\delta x\equiv x-z_0$.
$\epsilon\equiv S^{1/2}$ denotes the length of the short geodesic
section connecting $x$ to the worldline and normal to it, and $z_1$
is the point where this short geodesic intersects the worldline.
$u^{\alpha}$ and $u_1^{\alpha}$ denote the four-velocities at $z_0$
and $z_1$, respectively. $\hat u^{\alpha}$ and $\hat u_1^{\alpha}$
(the former not shown in the sketch) are vectors at $x$,
generated by parallelly-propagating the vectors $u^{\alpha}$ and
$u_1^{\alpha}$ along the short geodesic section from $z_0$ or
$z_1$, respectively, to $x$.
}
\label{figure}
\end{figure}

The direct part of the trace-reversed metric perturbation was obtained
by MST---see Eq.\ (2.27) of Ref.\ \cite{MST}. In the Appendix
we process the expression obtained by MST, and bring it to the form
\begin{equation} \label{II-10}
\bar h_{\beta\gamma}^{\rm dir}(x)=
4\mu\epsilon^{-1}\,\hat{u}^1_{\beta}\hat{u}^1_{\gamma}
+\epsilon^{-1}P^{(2)}_{\beta\gamma}(x,z_0).
\end{equation}
Here, $\epsilon$ is the spatial geodesic distance from the point $x$ to
the geodesic $z(\tau)$ (i.e., the length of the short geodesic connecting
$x$ to the worldline and normal to it), $z_1$ denotes the intersection of
this short normal geodesic with the worldline, and $\hat{u}^1_{\alpha}$
is the four-velocity parallelly-propagated (PP) from $z_1$ to $x$.
(See Fig.\ \ref{figure} for an illustration of the geometric setup
described here.)
The function $P^{(2)}_{\beta\gamma}$ is a regular function of $x$, of
order $\delta x^{2}$ (and higher orders), where $\delta x^{\mu}\equiv x^{\mu}-z_0^{\mu}$.
The explicit form of $P^{(2)}_{\beta\gamma}$ will not be needed in our
analysis.

For later convenience, we first re-express $\bar h_{\beta\gamma}^{\rm dir}$
in terms of the four-velocity PP from $z_0$ to $x$ (rather than from $z_1$
to $x$), which we denote by $\hat u_{\alpha}$ (or $\hat u^{\alpha}$).
Both $\hat u_{\alpha}$ and $\hat{u}^1_{\alpha}$ are regular functions of $x$,
and the difference between them is proportional to $\delta x^{2}$ (and to
the Riemann tensor). Absorbing this difference in the function
$P^{(2)}_{\beta\gamma}$, we may re-write the direct metric
perturbation as
\begin{equation} \label{II-20}
\bar{h}_{\beta\gamma}^{\rm dir}(x)
=4\mu S^{-1/2}\hat u_{\beta}\hat u_{\gamma}+
S^{-1/2}\tilde P^{(2)}_{\beta\gamma}(x,z_0),
\end{equation}
where $S\equiv\epsilon^2$, and the new function
$\tilde P^{(2)}_{\beta\gamma}$ has the same
features as $P^{(2)}_{\beta\gamma}$, namely, it is a regular function,
of order $\delta x^2$.

\subsection{Extending the tensor $k_0^{\alpha\beta\gamma\delta}$
off the worldline}

Given the above expression for the direct perturbation, the
direct force is constructed as a vector field through Eq.\ (\ref{I-75}).
In this equation, recall, $k^{\alpha\beta\gamma\delta}(x)$ is an
extension off the point $z_0$ of the tensor $k^{\alpha\beta\gamma\delta}_0$
defined at $z_0$ in terms of $u^{\alpha}(z_0)$ and
$g^{\alpha\beta}(z_0)$ [see Eq.\ (\ref{I-50})]. In our analysis we decompose
the components of the field $F^{\alpha}_{\rm dir}(x)$ in spherical harmonics.
Since this decomposition is nonlocal (it involves an integration over
the 2-sphere $r,t={\rm const}$), it will generally depend on the
extension of $k^{\alpha\beta\gamma\delta}_0$, which we now have to
specify.

A natural extension of $k^{\alpha\beta\gamma\delta}_0$
was prescribed by MST \cite{MST} (also MNS \cite{MNS}) by setting
in the right-hand side of Eq.\ (\ref{I-50}) $u^{\alpha}\to\hat u_1^{\alpha}(x)$
and $g^{\alpha\beta}\to g^{\alpha\beta}(x)$; namely, by PP the four
velocity $u^{\alpha}$ from $z_1$ to $x$, and assigning to the metric
function its actual value at $x$. However, for our analysis, we found it
useful to apply a different extension: one in which all (contravariant)
tensorial components of $k^{\alpha\beta\gamma\delta}(x)$
are assigned fixed values---the same values they have at $x\to z_0$:
\begin{equation} \label{II-40}
k^{\alpha\beta\gamma\delta}(x)\equiv k^{\alpha\beta\gamma\delta}_0.
\end{equation}
Note that this definition is coordinate-dependent; here we refer
to (contravariant components in) the {\em Schwarzschild coordinates}.
Throughout the rest of this
manuscript, $k^{\alpha\beta\gamma\delta}(x)$ will denote specifically
the extension defined in Eq.\ (\ref{II-40}), to which
we shall refer as the ``fixed components'' extension.
This extension turns out to be most convenient
for the numerical determination of the modes of the full force (recall
that the same choice of extension must be made for both the direct and
the full forces!).

\subsection{Constructing the direct force}

To analyze the direct force,
$F_{\rm dir}^{\alpha}(x)=\mu k^{\alpha\beta\gamma\delta}
\bar{h}^{\rm dir}_{\beta\gamma;\delta}$,
we first use Eq.\ (\ref{II-20}) to obtain
\begin{equation} \label{II-80}
\bar{h}^{\rm dir}_{\beta\gamma;\delta}=
-2\mu \epsilon^{-3}S_{,\delta}\hat u_{\beta}\hat u_{\gamma}
+4\mu\epsilon^{-1}(\hat u_{\beta;\delta}\hat u_{\gamma}+\hat u_{\beta}
\hat u_{\gamma;\delta})
-\epsilon^{-3}S_{,\delta}\tilde P^{(2)}_{\beta\gamma}/2
+\epsilon^{-1} P^{(1)}_{\beta\gamma\delta},
\end{equation}
where $P^{(1)}_{\beta\gamma\delta}\equiv \tilde
P^{(2)}_{\beta\gamma;\delta}$ is a regular function, of order $\delta x$ (and
higher orders).
Since $\hat u_{\alpha}$ is PP (from $z_0$ to $x$), its
covariant derivatives are proportional to $\delta x$.
Therefore, the second term in the above expression may be
absorbed in the fourth term: this merely amounts to modifying
the explicit form of $P^{(1)}_{\beta\gamma\delta}$.
Considering next the third, $\propto \epsilon^{-3}$ term in Eq.\
(\ref{II-80}), and recalling $S_{,\delta}\propto O(\delta x)$, we write this
term in the form
$\epsilon^{-3}P_{\beta\gamma\delta}^{(3)}$, where $P_{\beta\gamma\delta}^{(3)}$
is a regular quantity of $O(\delta x^3)$ (and higher orders). Absorbing then
the term $\epsilon^{-1}P_{\beta\gamma\delta}^{(1)}
=\epsilon^{-3}\left(\epsilon^{2}P_{\beta\gamma\delta}^{(1)}\right)$ in the term
$\epsilon^{-3}P_{\beta\gamma\delta}^{(3)}$ (which amounts to re-defining
$P_{\beta\gamma\delta }^{(3)}$), we finally write $\bar{h}^{\rm
dir}_{\beta\gamma;\delta}$ as
\begin{equation} \label{II-100}
\bar{h}^{\rm dir}_{\beta\gamma;\delta}=
-2 \mu\epsilon^{-3}S_{,\delta}\hat u_{\beta}\hat u_{\gamma}
+\epsilon^{-3}P_{\beta\gamma\delta}^{(3)}.
\end{equation}
Consequently, the direct force takes the form
\begin{equation} \label{II-110}
F_{\rm dir}^{\alpha}(x)=\mu^2\left(
-\frac{1}{2}K^{\alpha\delta}\epsilon^{-3}S_{,\delta}+\epsilon^{-3}
P^{\alpha}_{(3)}\right),
\end{equation}
where
\begin{equation} \label{II-120}
K^{\alpha\delta}\equiv 4k^{\alpha\beta\gamma\delta}
\hat u_{\beta}\hat u_{\gamma},
\end{equation}
and $\mu P^{\alpha}_{(3)}\equiv k^{\alpha\beta\gamma\delta}
P_{\beta\gamma\delta}^{(3)}$ (the factor $4$ is introduced for
later convenience).
Note that the quantity $P^{\alpha}_{(3)}$, which is regular
at $x=z_0$, generally contains also terms of order $\delta x^4$ and higher.
However, the contribution from such higher-order terms to
$F_{\rm dir}^{\alpha}$ vanishes at $\delta x=0$, and so these terms may be
ignored in our analysis. We shall indeed drop these higher-order terms, and
take $P^{\alpha}_{(3)}$ to be a polynomial in $\delta x$ of {\em homogeneous}
order $\delta x^{3}$.

The coefficients of the tensor $K^{\alpha\delta}$ are not constant,
as the field $\hat u_{\beta}$ is a PP field and not a field of constant
components. It will prove convenient to expand $K^{\alpha\delta}$ in
$\delta x$, and express it as
\begin{equation} \label{II-130}
K^{\alpha\delta}=K_0^{\alpha\delta}+K_1^{\alpha\delta}+
K_2^{\alpha\delta}+\dots,
\end{equation}
where $K_0^{\alpha\delta}[=K^{\alpha\delta}(x\to z_0)]$ is a field of
constant components, $K_1^{\alpha\delta}$ is proportional to $\delta x$,
and so on.
(Note that the terms $K_{n>0}^{\alpha\delta}$---unlike
$K_0^{\alpha\delta}$---depend on the extension.)
Considering now the first term in the expression for the direct force,
Eq.\ (\ref{II-110}), and recalling $S_{,\delta}\propto\delta x$, we
observe that the contribution from the term $K_3^{\alpha\delta}$
and higher order terms of $K^{\alpha\delta}$ to the direct force
vanishes at $x\to z_0$. We hence drop these terms.
In addition, we observe that the term
$K_2^{\alpha\delta}\epsilon^{-3}S_{,\delta}$ may be absorbed
in the term $\epsilon^{-3}P_{(3)}^{\alpha}$ of the direct force,
which merely amounts to re-defining $P_{(3)}^{\alpha}$.
Thus, the direct force takes the form
\begin{equation} \label{II-140}
F_{\rm dir}^{\alpha}=\mu^2\left(
\frac{1}{2}K_0^{\alpha\delta}\epsilon^{-3}S_{,\delta}+
\frac{1}{2}K_1^{\alpha\delta}\epsilon^{-3}S_{,\delta}+
\epsilon^{-3}P^{\alpha}_{(3)}\right).
\end{equation}

As in Paper I, we now expand $S$ in powers of $\delta x$, in the form
\begin{equation} \label{II-150}
S=S_0+S_1+S_2+\cdots,
\end{equation}
where $S_0$ is the leading order ($\propto\delta x^2$) term of $S$, $S_1$
is the correction term of homogeneous order $\delta x^3$, and so on.
In this work we will need only the explicit form of $S_0$:
\begin{equation} \label{II-152a}
S_0=(g_{\mu\nu}+u_{\mu}u_{\nu})\delta x^{\mu}\delta x^{\nu}.
\end{equation}
The factor $\epsilon^{-3}$ appearing in the last expression for the
direct force, Eq.\ (\ref{II-140}), is then expanded as
\begin{eqnarray} \label{II-155}
\epsilon^{-3}=S^{-3/2}&=&
S_0^{-3/2}-\frac{3}{2}S_0^{-5/2}S_1
+\left(\frac{15}{8}S_0^{-7/2} S_1^2-\frac{3}{2}S_0^{-5/2}S_2\right)
+\cdots \nonumber\\
&=&\epsilon_0^{-3}-\frac{3}{2}\epsilon_0^{-5}S_1
+\epsilon_0^{-7}\left(\frac{15}{8}S_1^2-\frac{3}{2}\epsilon_0^2 S_2\right)
+\cdots,
\end{eqnarray}
where $\epsilon_0\equiv S_0^{1/2}$. In this expansion, the first
term scales as $\delta x^{-3}$, the second as $\delta x^{-2}$,
and so on. The terms included in the three dots
scale as $\delta x^{0}$ or higher powers of $\delta x$.

Next we expand the direct force
in powers of $\delta x$, using the above expansions of $S$ and $\epsilon^{-3}$.
Based on Eq.\ (\ref{II-140}), this expansion takes the form
\begin{equation} \label{II-160}
F_{\rm dir}^{\alpha}=\mu^2\left(
\epsilon_0^{-3}P^{\alpha}_{(1)}+\epsilon_0^{-5}P^{\alpha}_{(4)}
+\epsilon_0^{-7}P^{\alpha}_{(7)}\right),
\end{equation}
in which $P^{\alpha}_{(n)}$ denote polynomials of homogeneous order $n$ in
$\delta x$, and where we have omitted higher-order terms that vanish
at $x\to z_0$. Notice that the term $\epsilon^{-3}P^{\alpha}_{(3)}$
of Eq.\ (\ref{II-140}) has been absorbed here in the term
$\epsilon_0^{-7}P^{\alpha}_{(7)}$ (with higher-order corrections
that vanish at $x\to z_0$ and are thus omitted). Also absorbed in
$\epsilon_0^{-7}P^{\alpha}_{(7)}$ are other terms like
$K_0^{\alpha\delta}\epsilon_0^{-3}S_{2,\delta}$,
$K_0^{\alpha\delta}\epsilon_0^{-5}S_1S_{1,\delta}$,
$K_1^{\alpha\delta}\epsilon_0^{-3}S_{1,\delta}$, etc.
The functions $P^{\alpha}_{(1)}$ and $P^{\alpha}_{(4)}$ are given
explicitly by
\begin{mathletters}\label{II-170}
\begin{equation} \label{II-170a}
P^{\alpha}_{(1)}=-\frac{1}{2}K_0^{\alpha\delta}S_{0,\delta},
\end{equation}
\begin{equation} \label{II-170b}
P^{\alpha}_{(4)}=-\frac{1}{2}K_0^{\alpha\delta}S_0S_{1,\delta}
+\frac{3}{4}K_0^{\alpha\delta}S_1S_{0,\delta}
-\frac{1}{2}K_1^{\alpha\delta}S_0S_{0,\delta}
\end{equation}
\end{mathletters}
(the explicit form of $P^{\alpha}_{(7)}$ will not be needed).
Note that the leading-order term of the direct force,
$\mu^2\epsilon_{0}^{-3}P^{\alpha}_{(1)}$, emerges
exclusively from the leading-order term $\propto K_0^{\alpha\delta}$
in Eq.\ (\ref{II-140}), whereas the next-order term, $\mu^2
\epsilon_{0}^{-5}P^{\alpha}_{(4)}$, is composed of contributions coming from
both terms $\propto K_0^{\alpha\delta}$ and $\propto K_1^{\alpha\delta}$.
The $\propto K_0^{\alpha\delta}$ contributions (and thus the entire
leading-order term) are all analogous to ones that occur in the
scalar model [see Eq.\ (23) of Paper I], whereas the
$\propto K_1^{\alpha\delta}$ contribution has no counterpart in the
scalar case considered therein. We also point out that, since $K_0^{\alpha\delta}$
does not depend on the extension of $k^{\alpha\beta\gamma\delta}$
(unlike $K_{n>0}^{\alpha\delta}$), one finds that the leading-order
term of $F_{\rm dir}^{\alpha}$ is extension-independent, whereas the
explicit form of the higher-oder terms does depend, in general,
on the choice of extension.

\section{Derivation of the regularization parameters: gravitational case}
\label{SecIII}

In principle, the derivation of the RP will now involve expanding the direct
force components $F_{\rm dir}^{\alpha}$ in scalar spherical harmonics, and
then taking the limit $x\to z_0$ [just as in the scalar case analysis---cf.\
Eq.\ (27) in Paper I]. This will yield the $l$-mode contribution to the direct
force, $F^{\alpha l}_{\rm dir}$, from which one may deduce the values of all RP.
However, at this point we may exploit the remarkable analogy between the
expression derived here for the gravitation direct force, Eq.\ (\ref{II-160}),
and the corresponding expression obtained in the scalar toy model [see Eq.\ (22)
of Paper I]: these expressions differ only in the explicit form of the three
coefficients $P^{\alpha}_{(1,4,7)}$. Conveniently, this analogy will now
allow us to base most of our analysis on results already obtained in Paper I.

We begin by recalling the expression obtained for the direct force in the
scalar case [see Eq.\ (27) in Paper I]\footnote{For later convenience,
we use here a re-definition of $F_{\rm \delta}^{{\rm (dir,sca)}}$, with the
factor $q^2$ omitted ($q$ is the scalar charge).}:
\begin{equation} \label{III-1}
F_{\rm \delta}^{{\rm (dir,sca)}}=
\epsilon_0^{-3}P_{\delta}^{(1,{\rm sca})}+
\epsilon_0^{-5}P_{\delta}^{(4,{\rm sca})}+
\epsilon_0^{-7}P_{\delta}^{(7,{\rm sca})},
\end{equation}
where $P_{\delta}^{(1,{\rm sca})}=-\frac{1}{2}S_{0,\delta}$,
$P_{\delta}^{(4,{\rm sca})}=-\frac{1}{2}S_0S_{1,\delta}
+\frac{3}{4}S_1S_{0,\delta}$, and $P_{\delta}^{(7,{\rm sca})}$
is a polynomial in $\delta x$, of homogeneous order $\delta x^7$,
whose explicit value will not be needed here. We use the label ``sca''
to distinguish quantities associated with the scalar case, from their
gravitational-case counterparts.

Comparing Eqs.\ (\ref{II-160}) and (\ref{III-1}), taking into account
also the explicit form of the coefficients $P_{\delta}^{(1,4,{\rm sca})}$
and $P_{\delta}^{(1,4)}$, we now express the gravitational direct force
as a sum of three terms, in the form
\begin{equation} \label{II-2}
F_{\rm dir}^{\alpha}=
\mu^2\left(F^{\alpha}_{1}+F^{\alpha}_{2}+F^{\alpha}_{3}\right),
\end{equation}
where
\begin{eqnarray} \label{II-3}
F_{1}^{\alpha}&\equiv&  K_0^{\alpha\delta}F^{\rm dir,sca}_{\delta},
\nonumber\\
F_{2}^{\alpha}&\equiv&
-\frac{1}{2}\epsilon_0^{-5}K_1^{\alpha\delta}S_0S_{0,\delta}=
-\frac{1}{2}\epsilon_0^{-3}K_1^{\alpha\delta}S_{0,\delta},
\nonumber\\
F_{3}^{\alpha}&\equiv& \epsilon_0^{-7}\left(P^{\alpha}_{(7)}
-K_0^{\alpha\delta} P_{\delta}^{(7,{\rm sca})}\right).
\end{eqnarray}
We proceed by considering separately the contributions to the RP from
each of the three terms $F_{1,2,3}^{\alpha}$.

\subsection{Contribution to the RP from the term $F^{\alpha}_1$}

Consider first the term $F_1^{\alpha}$ of the gravitational direct
force. This term is just the scalar direct force, contracted with
$K_0^{\alpha\delta}$---an array of constant coefficients
[recall $K_0^{\alpha\delta}\equiv K^{\alpha\delta}(\delta x\to 0)$,
where $K^{\alpha\delta}$ is the tensor defined in Eq.\ (\ref{II-120})].
Since the constant array $K_0^{\alpha\delta}$ does not interfere with the
multipole decomposition, one may immediately conclude that the contribution
from the term $F_1^{\alpha}$ to any of the RP, in the gravitational
case, would be precisely the same as in the scalar case---multiplied by
$K_0^{\alpha\delta}$. Denoting by $R_i^{\alpha}$ ($i=1,2,3$) the
contribution of the term $F_i^{\alpha}$ to any of the RP, we thus simply
have
\begin{equation} \label{III-20}
R_1^{\alpha}=K_0^{\alpha\delta}R_{\delta}^{\rm (sca)},
\end{equation}
where the scalar-case values $R_{\delta}^{\rm (sca)}$ are those given
explicitly in Paper I. (We have made here the obvious replacement
$q\to\mu$.) In particular, since $C_{\delta}^{\rm (sca)}=D_{\delta}^{\rm (sca)}=0$,
we find $C_1^{\alpha}=D_1^{\alpha}=0$.

\subsection{Contribution to the RP from the term $F^{\alpha}_2$}

We next consider the term $F_2^{\alpha}$ in Eq.\ (\ref{II-2}).
This term has the from $\propto \epsilon_0^{-5}P^{\alpha}_{(4)}$.
As shown in Paper I, in evaluating the contribution of this kind
of terms to the $l$-mode of the direct force at $z_0$, one is allowed
to take their limit $\delta t,\delta r\to 0$ {\em before} applying
the multipole decomposition. This is true regardless of the explicit
form of the polynomial $P^{\alpha}_{(4)}$. We hence proceed by
considering $F_2^{\alpha}(\delta t=\delta r=0)$; we show that this
quantity actually vanishes, even before applying the multipole
decomposition.

Examine the form of $F_2^{\alpha}$, as defined in Eq.\ (\ref{II-3}):
The quantity $K_1^{\alpha\delta}$, recall, is the first-order variation
of the tensor $K^{\alpha\delta}\equiv 4k^{\alpha\beta\gamma\delta}
\hat u_{\beta}\hat u_{\gamma}$ with respect to $\delta x$.
Recalling that $k^{\alpha\beta\gamma\delta}$ is a tensor of constant
components, we have $K_1^{\alpha\delta}=4k^{\alpha\beta\gamma\delta}
(\delta u_{\beta} u_{\gamma}+u_{\beta}\delta u_{\gamma})$,
where $\delta u_{\beta}$ is the first order variation in the PP
four-velocity $\hat u_{\beta}(x)$
(namely, $\delta u_{\beta}=\Gamma^{\lambda}_{\beta\rho}u_{\lambda}
\delta x^{\rho}$, with $\Gamma^{\lambda}_{\beta\rho}$ being the connection
coefficients at $z_0$).
Thus,
\begin{equation} \label{III-60}
F_2^{\alpha}\equiv -2\epsilon_0^{-3}k^{\alpha\beta\gamma\delta}
(\delta u_{\beta} u_{\gamma}+u_{\beta}\delta u_{\gamma})S_{0,\delta}.
\end{equation}

Consider now the explicit form of $k^{\alpha\beta\gamma\delta}$,
given in Eq.\ (\ref{I-50}). Three of the five terms of
$k^{\alpha\beta\gamma\delta}$ are proportional to $u^{\delta}$.
These three terms will contribute nothing to $F_2^{\alpha}$,
as
\[
u^{\delta}S_{0,\delta}=2u^{\delta}(g_{\mu\delta}+u_{\mu}u_{\delta})
\delta x^{\mu}=0.
\]

Consider next the first and fifth terms of $k^{\alpha\beta\gamma\delta}$,
proportional to $u^{\beta}u^{\gamma}$ and $g^{\beta\gamma}$, respectively.
Both terms yield contributions
to $F_2^{\alpha}$ which are proportional to $\delta u_{\beta}u^{\beta}$.
This quantity, in fact, vanishes for our orbital setup: To see that,
first recall $\hat u^{\beta}\hat u_{\beta}=u^{\beta}u_{\beta}=-1$,
as the length of the four velocity is preserved when PP. Then, observe
that the linear variation of this equality with respect to $\delta x$ yields
\begin{eqnarray} \label{III-63}
0=\delta(\hat u^{\beta}\hat u_{\beta})&=&
\delta u^{\beta}\hat u_{\beta}+\hat u^{\beta}\delta u_{\beta}
\nonumber\\
&=& 2\delta u_{\beta}u^{\beta}
+\delta g^{\alpha\beta}(x)\hat u_{\alpha} \hat u_{\beta}
\end{eqnarray}
(to linear order in $\delta x$), where
$\delta g^{\alpha\beta}(x)={g^{\alpha\beta}}_{,\gamma}(z_0)\delta x^{\gamma}$
is the linear variation in $g^{\alpha\beta}(x)$.
Since in our setup the trajectory is equatorial, and since
${g^{\alpha\beta}}_{,\theta}={g^{\alpha\beta}}_{,\varphi}=0$
at the equatorial plain, the linear variation $\delta g^{\alpha\beta}$
vanishes (recall that in considering $F_2^{\alpha}$ we reduce $\delta x$
to just $\delta\theta,\delta\varphi$).
Consequently, we obtain from Eq.\ (\ref{III-63}) $\delta u_{\beta}u^{\beta}$=0.


In conclusion of the above discussion, we find $F_2^{\alpha}=0$
(in the limit $\delta t=\delta r=0$). Hence, obviously, this term yields
no contribution to any of the RP:
\begin{equation} \label{III-65}
R_2^{\alpha}=0.
\end{equation}
Note that this result may no longer be valid when using $k$-extensions
other than the ``fixed components'' extension employed here:
Usually, there will arise additional terms in Eq.\ (\ref{III-60}),
corresponding to first-order variations of the tensor
$k^{\alpha\beta\gamma\delta}(x)$.
Also, notice that the result (\ref{III-65}) will generally not hold
when considering non-equatorial orbits, as the variation
$\delta g^{\alpha\beta}$ in Eq.\ (\ref{III-63}) will generally fail
to vanish.
In both cases (namely, a different $k$-extension, and/or a
non-equatorial orbit), our calculation would lead, in general,
to a nonvanishing contribution $R_2^{\alpha}$.

\subsection{Contribution to the RP from the term $F^{\alpha}_3$}

We finally turn to the term $F_3^{\alpha}$ in Eq.\ (\ref{II-2}).
Recalling that $K_0^{\alpha\delta}$ is just an array of constants,
we may write this term as
\begin{equation} \label{III-67}
F_{3}^{\alpha}=\epsilon_0^{-7}\tilde P^{\alpha}_{(7)},
\end{equation}
where $\tilde P^{\alpha}_{(7)}\equiv P^{\alpha}_{(7)}-
K_0^{\alpha\delta} P_{\delta}^{(7,{\rm sca})}$ is once again a polynomial
in $\delta x$, of order $\delta x^7$.

The contribution of the term $F_{3}^{\alpha}$ to the $l$-mode direct force
is obtained by carrying out the (Legendre) integration over a 2-sphere
$r=t={\rm const}$, and then taking the limits $\delta t,\delta r\to 0$.
As shown in Paper I, in evaluating the contribution of a term of the
form (\ref{III-67}) (regardless of the explicit form of
$\tilde P^{\alpha}_{(7)}$) one may interchange the integration and
the limits, and set $\delta t=\delta r=0$ before integrating
over the 2-sphere---just as with the term $F_{2}^{\alpha}$ considered above.
To carry out the Legendre integration, it proves especially
convenient---as in Paper I---to use a new set of spherical coordinates
$(\theta',\varphi')$, in which the particle is located at the polar axis,
$\theta'=0$.\footnote{
We should emphasize here that we do {\em not} regard $\theta',\varphi'$
as new spacetime coordinates---namely, all vectorial/tensorial quantities
are still taken with respect to the original coordinates $\theta,\varphi$.
That is, $\theta',\varphi'$ are merely used here as new variables
for implementing the Legendre integral. The same holds for the
coordinates $x,y$ introduced below.}
The contribution from $F_{3}^{\alpha}$ to the
$l$-mode direct force can then be expressed as
\begin{equation}\label{VIII-70}
\mu^2\frac{L}{2\pi}\int \hat \epsilon_0^{-7}\hat P^{\alpha}_{(7)}
P_l(\cos\theta')\,d(\cos\theta')d\varphi',
\end{equation}
where $P_l$ is the Legendre polynomial,
and $\hat \epsilon_0,\hat P^{\alpha}_{(7)}$ are the reductions of
$\epsilon_0,\tilde P^{\alpha}_{(7)}$, respectively, to $r=r_0$ and $t=t_0$.
(Note that, conveniently, in the $\theta',\varphi'$ system the contribution
to any $l$-mode at $x\to z_0$ comes only from the axially-symmetric, $m=0$ mode.)
From Eq.\ (\ref {II-152a}), recalling
$\epsilon_0=S_0^{1/2}$ and $u_{\theta}=0$, we obtain, explicitly,
\begin{equation}\label{VIII-71}
\hat{\epsilon}_{0}=
\left[r_0^2(\delta\theta^2+\delta\varphi^2)+
u_{\varphi}^2 \delta\varphi^2\right]^{1/2}.
\end{equation}

To implement the integral (\ref{VIII-70}), it proves
convenient, as in Paper I, to introduce Cartesian-like coordinates
$x,y$ on the 2-sphere, which we define here by
\begin{equation}\label{VIII-72}
x\equiv \theta'\cos\varphi',\quad\quad
y\equiv \theta'\sin\varphi'.
\end{equation}
Note $x=y=0$ at $z_0$, and hence we have simply $\delta x^x=x$ and
$\delta x^y=y$. It is simple to show that a choice of transformation
$(\theta,\varphi)\to(\theta',\varphi')$ can be made, such that the
coordinates $x,y$ would relate to the original coordinates
$\theta,\varphi$ through
\begin{equation}\label{VIII-74}
x=\delta\varphi+O(\delta x^2),\quad\quad
y=\delta\theta +O(\delta x^2).
\end{equation}
Expressed in terms of the new coordinates, the polynomial
$\hat P^{\alpha}_{(7)}(\delta\theta,\delta\varphi)$ in Eq.\
(\ref{VIII-70}) becomes $\bar P^{\alpha}_{(7)}(y,x)+O(\delta x^8)$,
where $\bar P^{\alpha}_{(7)}$ is a polynomial of homogeneous order $7$ in
$x,y$.
The contribution from the $O(\delta x^8)$ corrections to the direct
force vanishes at $x\to z_0$, and can therefore be omitted.
From Eq.\ (\ref{VIII-71}) we also get
\begin{equation}\label{VIII-75}
\hat{\epsilon}_{0}=\bar\epsilon_0 (x,y)+O(\delta x^2),
\end{equation}
where $\bar\epsilon_0\equiv
\left[r_0^2(x^2+y^2)+u_{\varphi}^2 x^2\right]^{1/2}$.
Again, only the leading-order term here contributes to
the direct force at $x\to z_0$, and we are allowed to drop the
$O(\delta x^2)$ correction.
Hence, as far as the calculation of the RP is concerned, we may
express the contribution from the term $F_{3}^{\alpha}$ to the $l$-mode
direct force, Eq.\ (\ref{VIII-70}), as
\begin{equation}\label{VIII-81}
\mu^2\frac{L}{2\pi}\int \bar\epsilon_0^{\,-7}\bar P^{\alpha}_{(7)}(x,y)
P_l(\cos\theta')\,dxdy.
\end{equation}
Note that the Jacobian of the transformation
$(\cos\theta',\varphi')\to (x,y)$, which is actually given by $1+O(\delta x^2)$,
has been set here to just $1$: The higher-order corrections are once again
omitted, as they vanish at $x\to z_0$.

Examine now the integral in Eq.\ (\ref{VIII-81}):
$\bar{\epsilon}_{0}$ is an even function of
both $x$ and $y$. So is the function $\cos\theta'$. However, each of the
possible individual terms in the polynomial $\bar P^{\alpha}_{(7)}$
(like $\propto xy^6$, or $\propto x^4y^3$, for instance)
is necessarily an {\em odd} function of either $x$ or
$y$. Consequently, we observe that the entire integrand
in Eq.\ (\ref{VIII-81}) is odd in either $x$ or
$y$. Therefore, obviously, the integration over the 2-sphere
would vanish. As a consequence, no contribution to the RP will arise
from the term $F_3^{\alpha}$:
\begin{equation} \label{III-85}
R_3^{\alpha}=0.
\end{equation}
Notice that this last result is valid for any (sufficiently regular)
$k$-extension. A modification of the extension
would only affect the explicit form of the polynomials
$\bar P^{\alpha}_{(7)}$, but would not alter the odd parity structure
of the integrand in Eq.\ (\ref{VIII-81}).

\subsection{Summary: RP values in the gravitational case.}

Let us now collect the above results: We have found that neither
of the terms $F_2^{\alpha}$ and $F_3^{\alpha}$ contributes to the
$l$-mode direct force. The sole contribution to the RP comes from
the term $F_1^{\alpha}$---this contribution is given in Eq.\
(\ref{III-20}). The RP in the gravitational case are therefore given by
\begin{equation} \label{III-200}
R^{\alpha}=K_0^{\alpha\delta}R_{\delta}^{\rm (sca)},
\end{equation}
where, recall, $R^{\alpha}$ stands for any of the RP, and the
scalar-case values $R_{\delta}^{\rm (sca)}$ are those given explicitly
in Paper I.
We now need only to provide the explicit form of $K_0^{\alpha\delta}$:
Recalling $K_0^{\alpha\delta}=K^{\alpha\delta}(x\to z_0)$,
one easily gets from Eqs.\ (\ref{II-120}) and (\ref{I-50})
\begin{equation} \label{III-30}
K_0^{\alpha\delta}=
g^{\alpha\delta}+u^{\alpha}u^{\delta},
\end{equation}
where, recall, $u^{\alpha}$ and $g^{\alpha\delta}$ denote the values of
these quantities at $z_0$. Note that $K_0^{\alpha\delta}$ is just the
spatial projection operator at $z_0$, namely,
$\left(K_0^{\alpha\delta}V_{\delta}\right)u_{\alpha}=0$
for any vector $V_{\delta}$.

Let us finally write Eq.\ (\ref{III-200}) more explicitly:
First, recalling (see Paper I) that the scalar parameter
$A_{\delta}^{\rm (sca)}$ has no component tangent to $u^{\alpha}$
(i.e., $u^{\delta}A_{\delta}^{\rm (sca)}=0$), we simply obtain
\begin{equation} \label{III-40}
A^{\alpha}=A^{\alpha}_{\rm (sca)}
\end{equation}
(with the obvious substitution $q\to\mu$).
Unlike the situation with the parameter $A^{\alpha}$, the quantity
$u^{\delta}B_{\delta}^{\rm (sca)}$ does not vanish [see Eq.\ (85) in
Paper I], and we leave the expression for $B^{\alpha}$ in the form
\begin{equation} \label{III-90}
B^{\alpha}=K_0^{\alpha\delta}B_{\delta}^{\rm (sca)}
\end{equation}
(again, with $q\to\mu$). Finally, as $C_{\delta}^{\rm (sca)}
=D_{\delta}^{\rm (sca)}=0$, we shall have, in the gravitational case alike,
\begin{equation} \label{III-5}
C^{\alpha}=D^{\alpha}=0.
\end{equation}


\section{Derivation of the regularization parameters: electromagnetic case}
\label{SecV}

In this section we consider the {\em electromagnetic} self force acting
on an electrically charged particle: we prescribe the mode sum
scheme in this case, and construct all required RP for an arbitrary
(equatorial)
geodesic orbit in Schwarzschild spacetime. The same analytic calculation used
for deriving the gravitational-case RP will prove directly applicable
also to the electromagnetic case, with only minor adaptations required.

We shall consider a particle carrying an electric charge $e$ (with $|e|\ll M$),
and assume the same orbital configuration as in the gravitational case
(namely, the particle is taken to move along an equatorial orbit, which in the
limit $e\to 0$ becomes a geodesic). We shall also maintain here the notation
for the various
quantities $z_0$, $x$, $\epsilon$, $S$, $\hat u_{\alpha}$, and so on.
We shall denote by $\phi_{\alpha}(x)$ the vector potential associated with the
``full'' electromagnetic field induced by the particle. In this section we
ignore the gravitational self force.

A formal expression for the electromagnetic self force in curved spacetime was
obtained long ago by DeWitt and Brehme \cite{DB} (and was reproduced recently by
QW \cite{QW} using a different method). For a geodesic in vacuum spacetime, the
electromagnetic self force is obtained from an electromagnetic ``tail'' force,
just as in the gravitational case:
\begin{equation}\label{V-2}
F^{\alpha {\rm (EM)}}_{\rm self}=
\lim_{x\to z_0} F^{\alpha{\rm (EM)}}_{\rm tail}(x),
\end{equation}
where hereafter we use the label ``EM'' to signify quantities associated
with the electromagnetic case. The formal construction of the vector field
$F^{\alpha{\rm (EM)}}_{\rm tail}(x)$ is described in \cite{DB,QW}. 
Like in the gravitational case, the electromagnetic tail force can be written
as the difference between a ``full'' force and a ``direct'' force---just as
in Eq.\ (\ref{I-60}). In the electromagnetic case, these two vector fields
are given by \cite{MNS}
\begin{equation}\label{V-10}
F^{\alpha{\rm (EM)}}_{\rm full}(x)=
e k^{\alpha\beta\gamma}\phi_{\beta;\gamma}, \quad\quad
F^{\alpha{\rm (EM)}}_{\rm dir}=
e k^{\alpha\beta\gamma}\phi^{\rm dir}_{\beta;\gamma},
\end{equation}
where $\phi^{\rm dir}_{\beta}(x)$ is the ``direct'' part of the vector
potential (given explicitly below), and $k^{\alpha\beta\gamma}(x)$ is a
(sufficiently regular) extension of the tensor
\begin{equation}\label{V-20}
k_0^{\alpha\beta\gamma}\equiv
g^{\alpha\gamma}u^{\beta}-g^{\alpha\beta}u^{\gamma},
\end{equation}
defined at $z_0$.
As in the gravitational case, we shall adopt here the ``fixed components''
extension, defined (in Schwarzschild coordinates) through
$k^{\alpha\beta\gamma}(x)\equiv k_0^{\alpha\beta\gamma}$.

The mode sum prescription for the electromagnetic self force is completely
analogous to the one prescribed in the gravitational and scalar cases:
Given the (scalar harmonic) $l$-modes $F^{\alpha l{\rm(EM)}}_{\rm full}$
of the electromagnetic full force, the electromagnetic self force is
constructed through
\begin{equation}\label{V-25}
F^{\alpha{\rm(EM)}}_{\rm self}=\sum_{l=0}^{\infty}\left[\lim_{x\to z_0}
F^{\alpha l{\rm(EM)}}_{\rm full}(x)-A^{\alpha{\rm(EM)}}L-B^{\alpha{\rm(EM)}}
-C^{\alpha{\rm(EM)}}/L\right]-D^{\alpha{\rm(EM)}},
\end{equation}
where the various electromagnetic-case RP are to be obtained, again, by
analyzing the multipole modes of the direct force. The rest of this
section is devoted to calculating these electromagnetic RP.

As in the gravitational case, our starting point would be the expression
for the direct part of the particle's field---this time the direct part of
the vector potential---as obtained by MNS [see Eq.\ (B3) of Ref.\ \cite{MNS}].
In precisely the same manner as in the gravitational case, this expression
can be brought to the form
\begin{equation}\label{V-30}
\phi_{\beta}^{\rm dir}(x)=e S^{-1/2}\hat
u_{\beta}+S^{-1/2}P^{(2)}_{\beta}(x)
\end{equation}
[in analogy with Eq.\ (\ref{II-20})], where $P^{(2)}_{\beta}$ is a (regular)
function of $O(\delta x^2)$. The derivatives of the direct vector potential
then take the form
\begin{equation}\label{V-40}
\phi_{\beta;\gamma}^{\rm dir}=
-(e/2)S^{-3/2}S_{,\gamma}\hat u_{\beta}
+e S^{-1/2}\hat u_{\beta;\gamma}
-S^{-3/2}S_{,\gamma}P^{(2)}_{\beta}/2
+S^{-1/2}P^{(2)}_{\beta;\gamma},
\end{equation}
which [as in obtaining Eq.\ (\ref{II-100})] can be put into the form
\begin{equation}\label{V-50}
\phi_{\beta;\gamma}^{\rm dir}=
-\frac{1}{2}e\epsilon^{-3}S_{,\gamma}\hat u_{\beta}
+\epsilon^{-3} P^{(3)}_{\beta\gamma},
\end{equation}
with $P^{(3)}_{\beta\gamma}$ being a (regular) function of order $\delta x^3$
(and higher orders). Consequently, the direct electromagnetic force takes
precisely the same form as in Eq.\ (\ref{II-110}),
\begin{equation} \label{V-60}
F_{\rm dir}^{\alpha{\rm (EM)}}(x)=e^2\left(
-\frac{1}{2}K_{\rm EM}^{\alpha\delta}\epsilon^{-3}S_{,\delta}
+\epsilon^{-3} P^{\alpha}_{(3)}\right),
\end{equation}
where this time
\begin{equation} \label{V-70}
K_{\rm (EM)}^{\alpha\delta}\equiv k^{\alpha\beta\delta}\hat u_{\beta}
\end{equation}
and $eP^{\alpha}_{(3)}\equiv k^{\alpha\beta\gamma}P^{(3)}_{\beta\gamma}$.
Again, we may drop all terms of $P^{\alpha}_{(3)}$ which are of order
$\delta x^4$ and higher (as they do not contribute to the direct force at
$x\to z_0$), and take $P^{\alpha}_{(3)}$ to be of homogeneous order
$\delta x^3$.

Thanks to the complete analogy between the forms of the electromagnetic
and gravitational direct forces [compare Eqs.\ (\ref{V-60}) and (\ref{II-110})],
our analysis now proceeds precisely as in the gravitational case:
We expand $K_{\rm (EM)}^{\alpha\delta}$ and $S$ in $\delta x$, as in Eqs.\
(\ref{II-130}) and (\ref{II-150}), and consequently write the direct
force as a sum of three terms---in precise analogy with Eqs.\ (\ref{II-160})
and (\ref{II-2}):
\begin{eqnarray} \label{V-73}
F_{\rm dir}^{\alpha{\rm(EM)}}&=&e^2\left(
\epsilon_0^{-3}P^{\alpha{\rm(EM)}}_{(1)}+\epsilon_0^{-5}P^{\alpha{\rm(EM)}}_{(4)}
+\epsilon_0^{-7}P^{\alpha{\rm(EM)}}_{(7)}\right) \nonumber\\
&\equiv&
e^2\left(F^{\alpha{\rm(EM)}}_{1}+F^{\alpha{\rm(EM)}}_{2}
+F^{\alpha{\rm(EM)}}_{3}
\right),
\end{eqnarray}
where $F^{\alpha{\rm(EM)}}_{1,2,3}$ are defined in
Eq.\ (\ref{II-3}), with the replacements
$K_n^{\alpha\delta}\to K_{n\rm (EM)}^{\alpha\delta}$ and
$P^{\alpha}_{(7)}\to P^{\alpha{\rm(EM)}}_{(7)}$.
The only point at which our current analysis differs from the gravitational
case is in the explicit values taken by the various coefficients
$K_n^{\alpha\delta}$ (and consequently in the explicit values of the terms
$P^{\alpha}_{(n)}$).

Consider first the contribution to the $l$-mode direct force coming from
the term $F^{\alpha{\rm(EM)}}_{1}$:
In a complete analogy with Eq.\ (\ref{III-20}), we obtain
\begin{equation} \label{III-77}
R^{\alpha}_{1\rm (EM)}=K_{0{\rm (EM)}}^{\alpha\delta}R_{\delta}^{\rm (sca)}
(q\to e),
\end{equation}
where $R_{i{\rm(EM)}}^{\alpha}$ ($i=1,2,3$) stands for the contribution of the
term $F^{\alpha{\rm(EM)}}_{i}$ to any of the RP,
and the array of constant coefficients
$K_{0{\rm (EM)}}^{\alpha\delta}\equiv
K_{\rm (EM)}^{\alpha\delta}(x\to z_0)$ is now given by
\begin{equation} \label{V-80}
K_{0{\rm (EM)}}^{\alpha\delta}=-(g^{\alpha\delta}+u^{\alpha}u^{\delta}).
\end{equation}
Noticing $K_{0{\rm (EM)}}^{\alpha\delta}=-K_{0}^{\alpha\delta}$
[compare Eq.\ (\ref{V-80}) to Eq.\ (\ref{III-30})] and recalling
Eqs.\ (\ref{III-20}) and (\ref{III-200}), we then conclude
\begin{equation} \label{V-90}
R^{\alpha}_{1\rm (EM)}=-R^{\alpha}_{1\rm (grav)}(\mu\to e)=
-R^{\alpha}_{\rm (grav)} (\mu\to e),
\end{equation}
where hereafter we use the label ``grav'' to
signify the gravitational-case values.

Next, consider the term
$F^{\alpha{\rm(EM)}}_{2}\equiv -\frac{1}{2}\epsilon_0^{-3}
K_{1{\rm (EM)}}^{\alpha\delta} S_{0,\delta}$.
Here, the coefficient $K_{1{\rm(EM)}}^{\alpha\delta}$ (the first order
correction in $K_{\rm(EM)}^{\alpha\delta}$) is given by
\begin{equation} \label{V-91}
K_{1{\rm(EM)}}^{\alpha\delta}=k^{\alpha\beta\delta}\delta u_{\beta}.
\end{equation}
As in the gravitational case, it is easy to show that
$F^{\alpha{\rm(EM)}}_{2}$ (evaluated at $\delta t=\delta r=0$)
actually vanishes, even before taking its multipole decomposition:
From Eq.\ (\ref{V-20}) we observe that $k^{\alpha\beta\delta}$
is composed of two terms, one proportional to $u^{\beta}$ and the other
proportional to $u^{\delta}$. The $\propto u^{\beta}$ term contributes nothing
to $K_{1{\rm(EM)}}^{\alpha\delta}$, since $u^{\beta}\delta u_{\beta}=0$ (as
explained when discussing the gravitational case). The $\propto u^{\delta}$
term will yield a zero contribution as well, by virtue of $S_{0,\delta}u^{\delta}=0$.
We thus find that in the electromagnetic case---just as in the
gravitational case---the term $F^{\alpha{\rm(EM)}}_{2}$ contributes
nothing to the RP, namely, $R^{\alpha}_{2\rm (EM)}=0$.

As to the last term in the electromagnetic direct force,
$F^{\alpha{\rm(EM)}}_{3}$: Using the same parity considerations as in
the gravitational case, one shows that the contribution from this term
to any of the RP will vanish, namely, $R^{\alpha}_{3\rm (EM)}=0$.
This vanishing is irrespective of the explicit form of the polynomial
$P^{\alpha{\rm(EM)}}_{(7)}$ in Eq.\ (\ref{V-73}).

In conclusion, thus, we find that the sole contribution to the RP
in the electromagnetic case comes from the term $F^{\alpha{\rm(EM)}}_{1}$,
Eq.\ (\ref{V-90}). We hence obtain
\begin{equation} \label{V-99}
R^{\alpha}_{\rm (EM)}=-R^{\alpha}_{\rm (grav)}(\mu\to e).
\end{equation}

\section{Summary and concluding remarks}\label{SecVI}

Let us summarize our mode-sum prescription for constructing the
gravitational and electromagnetic self forces. We start with the
gravitational case:
\begin{enumerate}
\item
For a given trajectory, compute the tensor-harmonic modes of the
metric perturbation, $\bar{h}^{(i)l'm}(r,t)$, by numerically
integrating the separable field equations (e.g., in the harmonic gauge
\cite{MSRS-grav}).
\item
Given $\bar{h}^{(i)l'm}(r,t)$, construct the full modes
$F^{\alpha l}_{\rm full}$ at the particle's location.
This is done by applying the operator in Eq.\ (\ref{I-70}) to
$\bar{h}^{(i)l'm}(r,t)$, using the ``fixed components'' extension described
above, and then expanding the resultant field into scalar spherical
harmonics (and summing over $il'm$ for a given $l$).
This procedure is implemented in Refs.\ \cite{BL-radial,BL-circular}.
\item
Use Eqs.\ (\ref{RP}) below [along with Eqs.\ (83) of Paper I] to obtain the
RP values corresponding to the trajectory under consideration.
\item
Finally, apply the mode-sum formula, Eq.\ (\ref{I-90}).
\end{enumerate}
This prescription is now being implemented by Barack and Lousto
for radial \cite{BL-radial} and circular \cite{BL-circular} orbits in
Schwarzschild spacetime.

The prescription for constructing the electromagnetic self force
is similar: First, one has to compute the vector-harmonic modes
of the (full-field) vector potential for the given orbital configuration. Then,
one constructs the full-force modes $F^{\alpha l{\rm(EM)}}_{\rm full}$---this
construction is carried out by applying Eq.\ (\ref{V-10})
to each of the full-field vector-harmonic modes, and then decomposing
each of these modes in scalar spherical harmonics.
Finally, one applies the mode-sum formula (\ref{V-25}), with the
electromagnetic RP values given in Eq.\ (\ref{RP}) below.

The values of the RP in the gravitational and electromagnetic cases
are summarized as follows:
\begin{mathletters}\label{RP}
\begin{equation}\label{RPA}
A_{\alpha}^{\rm(grav)}=-A_{\alpha}^{\rm(EM)}=A_{\alpha}^{\rm(sca)},
\end{equation}
\begin{equation}\label{RPB}
B_{\alpha}^{\rm(grav)}=-B_{\alpha}^{\rm(EM)}=\left(\delta_{\alpha}^{\beta}+
u_{\alpha}u^{\beta}\right)B_{\beta}^{\rm (sca)},
\end{equation}
\begin{equation}\label{RPCD}
C^{\rm (grav,EM)}_{\alpha}=D^{\rm (grav,EM)}_{\alpha}=0
\end{equation}
\end{mathletters}
(with the obvious replacements $q\to\mu$ or $q\to e$),
where the quantities labeled ``sca'' are the scalar-field parameters
given explicitly in Eqs.\ (83) of paper I.
The gravitational RP were calculated previously in a different
method, using the $l$-mode Green's function expansion technique
\cite{MSRS-grav,BL-radial}, in the special case of radial orbits.
The results agree with the values (\ref{RP}).

As we discussed above, there is a certain ambiguity in the values of the
RP, which arises from the freedom in choosing the extension of the
tensor $k$ off the evaluation point $z_0$ (the choice of this extension
affects, of course,
the multipole decomposition of the force). However, our mode-sum scheme
produces no ambiguity in the eventual value of the self force---one only
has to make sure that the full-force modes $F^{\alpha l}_{\rm full}$
in Eq.\ (\ref{I-90}) [or in Eq.\ (\ref{V-25})] are calculated using the
same extension as the one used in calculating the RP. It is thus essential to
recall here that the RP values summarized above are those referring to the
``fixed components'' extension of the tensors $k^{\alpha\beta\gamma\delta}$
or $k^{\alpha\beta\gamma}$ (expressed in Schwarzschild coordinates).
This extension is most easily applicable in the numerical computation of
the full-force modes \cite{BL-radial,BL-circular}.


Based on our above analysis, we may phrase the following general
statements concerning the extension-dependence of the RP
in the electromagnetic and gravitational cases:
(i) The RP $A^{\alpha}$, $C^{\alpha}$, and $D^{\alpha}$ are
{\em insensitive} to the extension of $k^{\alpha\beta\gamma\delta}$
(provided it is regular enough).
(ii) The value of $B^{\alpha}$ {\em does} depend, in general, on the choice
of extension; however, all sufficiently regular extensions which differ
from the ``fixed components'' extension $k^{\alpha\beta\gamma\delta}$
by an amount of only $O(\delta x^2)$ will
admit the same value of $B^{\alpha}$---the one given in Eq.\ (\ref{RPB}).
It is interesting to refer here to MNS's analysis \cite{MNS}, in which
a different extension has been employed:
MNS extended the tensor $k^{\alpha\beta\gamma\delta}$ by PP the four
velocity from $z_0$ to $x$ and just assigning to $g^{\alpha\beta}$ the
actual value it has at $x$. Interestingly, within this extension
[differing from the ``fixed components'' extension already at
$O(\delta x)$], all RP attain precisely the same values as in the scalar
case \cite{Letter} (except that in the electromagnetic case all
RP are to be multiplied by $-1$).

Finally, it is important to remind that the gravitational self force is
a {\em
gauge-dependent} notion, as discussed in Ref.\ \cite{gauge}. The prescription
described in this manuscript applies to the self force associated with the {\em
harmonic gauge} (in which the original MST/QW scheme has been formulated). It
also applies, with the same RP values, to any other gauge related to the
harmonic gauge through a regular gauge transformation \cite{gauge}. However,
for other, non regular gauges, the mode-sum scheme is not guaranteed to be
valid in its present form.
A method for overcoming this gauge problem has been sketched in \cite{gauge},
and is currently being implemented for circular orbits in Schwarzschild
\cite{BL-circular}. A different strategy (applicable in the Schwarzschild case)
would be to calculate the self force directly in the harmonic gauge \cite{SNS}.

\section*{acknowledgements}
We are grateful to Lior Burko, Yasushi Mino, Hiroyuki Nakano, and
Misao Sasaki for interesting discussions and stimulating interaction.
L.B.\ was supported by a Marie Curie Fellowship of the European Community
program IHP-MCIF-99-1 under contract number HPMF-CT-2000-00851.

\appendix

\section{direct metric perturbation}\label{AppA}

In this appendix we obtain Eq.\ (\ref{II-10}) for the direct metric
perturbation, by processing the expression given by MST in
Ref.\ \cite{MST}.

By considering the Hadamard expansion of the (full) metric perturbation,
MST obtained the following expression for the (retarded) trace-reversed
perturbation [see Eq.\ (2.27) of Ref.\ \cite{MST}]:
\begin{equation} \label{AppA-10}
\bar h_{\beta\gamma}(x)= 2\mu\left[
\frac{2{\beta}(x,z_0)}{\epsilon}\,\hat{u}^1_{\beta}\hat{u}^1_{\gamma}
-u_1^{\sigma} S^{;\lambda}u_1^{\rho}R^1_{\sigma\lambda\rho(\hat\beta}
\hat u^1_{\gamma)}
+2\epsilon R^1_{\hat\beta\lambda\hat\gamma\sigma}u_1^{\lambda}u_1^{\sigma}
\right] +\text{tail term} +O(\epsilon^2).
\end{equation}
Here we use the notation of our Sec.\ \ref{SecII} (see Fig.\ \ref{figure}),
namely: $x$ is a point in the neighborhood of the force evaluation
point $z_0$; $\epsilon\equiv S^{1/2}$ is the length of the short
geodesic section connecting $x$ to the worldline and normal to it;
$z_1$ denotes the intersection of this geodesic with the worldline;
$u^{\alpha}$ and $u_1^{\alpha}$ (or $u^1_{\alpha}$) denote the
four-velocities at $z_0$ and $z_1$, respectively; and $\hat u^{\alpha}$
and $\hat u_1^{\alpha}$ (or $\hat u^1_{\alpha}$) are their PP to $x$.
In addition, parentesized indices denote symmetrization,
and $R^1_{\alpha\beta\gamma\delta}$ represents the Riemann tensor PP from
$z_1$ to $x$ with respect to any of its indices carrying the hat sign.
The ``tail term'' represents a non-local contribution to the full
perturbation, with its form given explicitly in \cite{MST}.
The function $\beta$ (denoted $\kappa^{-1}$ in \cite{MST}) is a regular
function satisfying $\beta=1+O(\delta x^2)$ [see Eq.\ (A14) therein].
Note that the correction term proportional to the 4-acceleration
in Eq.\ (2.27) of Ref.\ \cite{MST} can be omitted, since, for geodesic
orbits, it contributes to the self force only at order higher than $O(\mu^2)$.
For the same reason, we omit here the $O(\tau_r^{-1}\epsilon)$ term indicated
therein. Finally, notice the notational change $\sigma\to S/2$.

The direct part of the metric perturbation is now taken as
the difference between the full perturbation given in Eq.\ (\ref{AppA-10}),
and the tail term. The terms included in $O(\epsilon^2)$ do not contribute
to the direct perturbation at the limit $x\to z_0$ ($\epsilon\to 0$);
nor do they contribute to the direct force, whose construction involves
only first-order derivatives of $\bar h_{\alpha\beta}(x)$. We thus re-define
the direct perturbation by ignoring these $O(\epsilon^2)$ terms, which leaves
us with only the three terms in the squared brackets, scaling as
$\epsilon^{-1}$, $\epsilon^1$, and $\epsilon^1$, respectively.

Consider now the second and third terms in the squared brackets:
First, note that the (coordinate components of the) two vectors $u_1^{\alpha}$
and $u^{\alpha}$ differ only at $O(\delta x)$. Hence [recalling
$\epsilon,S^{;\lambda}\propto O(\delta x)$], this difference contributes
only to $O(\delta x^2)$ in Eq.\ (\ref{AppA-10}). We may thus ignore this
correction, and just replace $u_1^{\alpha}$ with $u^{\alpha}$ in the second and
third terms in the squared brackets. Likewise, we replace $\hat u_1^{\alpha}$
with $\hat u^{\alpha}$ in the second term. Similarly, we may ignore the
$O(\delta x)$ difference between $R^1_{\alpha\beta\gamma\delta}$ and
$R_{\alpha\beta\gamma\delta}$ (the latter denoting the coordinate-value of the
Riemann tensor at $z_0$) as it contributes only to $O(\delta x^2)$ in Eq.\
(\ref{AppA-10}). The direct metric perturbation thus takes the form
\begin{equation} \label{AppA-20}
\bar h_{\beta\gamma}^{\rm dir}(x)= 2\mu\left[
\frac{2{\beta}(x,z_0)}{\epsilon}\,\hat{u}^1_{\beta}\hat{u}^1_{\gamma}
-u^{\sigma} S^{;\lambda}u^{\rho}R_{\sigma\lambda\rho(\hat\beta}\hat u_{\gamma)}
+2\epsilon R_{\hat\beta\lambda\hat\gamma\sigma}u^{\lambda}u^{\sigma}
\right].
\end{equation}

Examine now more closely the second term in Eq.\ (\ref{AppA-20}):
Since $S^{;\lambda}\propto O(\delta x)$, the only contribution to the direct
force which does not vanish at $x\to z_0$ arises from differentiating
$S^{;\lambda}$. Recalling Eq.\ (\ref{II-152a}), we have
\[
S^{;\lambda}_{\ ;\delta}=2(\delta^{\lambda}_{\delta}
+u^{\lambda}u_{\delta})
+O(\delta x).
\]
Note that $S^{;\lambda}_{\ ;\delta}u^{\delta}=0$ (at $x\to z_0$).
Note also that the second term in Eq.\ (\ref{AppA-20}) (unlike the
other two terms) is perfectly regular at $x=z_0$. This allows us to
evaluate its contribution directly at $z_0$, which we do by just ``removing''
the hat symbols from $R_{\sigma\lambda\rho\hat\beta}$ and $\hat u_{\gamma}$.
Recalling Eq.\ (\ref{I-75}), the contribution from this term to the
direct force at $z_0$ then reads
\[
-2\mu^2 k^{\alpha\beta\gamma\delta}u^{\sigma}S^{;\lambda}_{\ ;\delta}
u^{\rho}R_{\sigma\lambda\rho(\beta}u_{\gamma)}
\]
(evaluates at $x=z_0$).
Examining the form of the tensor $k^{\alpha\beta\gamma\delta}$,
given in Eq.\ (\ref{I-50}), we observe that three of its five terms
are proportional to $u^{\delta}$, and thus vanish when contracted
with $S^{;\lambda}_{\ ;\delta}$. The first term of
$k^{\alpha\beta\gamma\delta}$ is proportional to $u^{\beta}u^{\gamma}$,
and thus yields a vanishing contribution when contracted
with either $u^{\rho}R_{\sigma\lambda\rho\beta}$ or
$u^{\rho}R_{\sigma\lambda\rho\gamma}$.
Likewise, the last term of $k^{\alpha\beta\gamma\delta}$, proportional
to $g^{\gamma\beta}$, is found to vanish when contracted with either
$u^{\rho}R_{\sigma\lambda\rho\beta}u_{\gamma}$ or
$u^{\rho}R_{\sigma\lambda\rho\gamma}u_{\beta}$.
We conclude that the contribution of this regular term to the
self force vanishes---even before taking the harmonic decomposition.

Finally, consider the third term in Eq.\ (\ref{AppA-20}).
Noticing that this term has the form $\epsilon\times$ (a regular
function of $x$), we may write it as
\begin{equation} \label{AppA-30}
\epsilon^{-1}P^{(2)}_{\beta\gamma}(x,z_0),
\end{equation}
where $P^{(2)}_{\beta\gamma}$ is a certain regular function, of order $\delta x^2$.
(This form will be convenient for our analysis in Sec.\ II.)
We further notice that the terms of $O(\delta x^2)$ included in the
function $\beta$ contribute to $\bar h_{\beta\gamma}$ an amount of precisely
the form (\ref{AppA-30}). We may thus absorb this contribution in the
contribution (\ref{AppA-30}) coming from the third term, and replace
the function $\beta$ with just $1$. The explicit form of the regular
function $P^{2}_{\beta\gamma}(x)$ will not be needed in our analysis.

In conclusion, we find that the direct part of the metric perturbation
is effectively given by Eq.\ (\ref{II-10}). This expression is used
as a starting point for the analysis in this paper.


\end{document}